\documentclass[twocolumn,english,aps,prd,showpacs,superscriptaddress]{revtex4-1}
\pdfoutput=1
\usepackage[latin9]{inputenc}
\usepackage{amsmath}
\usepackage{amsthm}
\usepackage{amssymb}
\usepackage{esint}

\makeatletter
 \usepackage{slashed}
\newcommand{\Lagr}{\mathcal{L}}

\usepackage{feyn}
\usepackage{feynmp}
\newcommand{\Tr}{\mathrm{Tr}}
\newcommand{\F}{\mathrm{F}}
\makeatother

\usepackage{babel}
\begin{document}

\title{Quantum Valley Hall Effect in Massive Dirac Systems Coupled to a
Scalar Field}

\author{S.H. Kooi}

\affiliation{Institute for Theoretical Physics, Centre for Extreme Matter and
Emergent Phenomena,\\ Utrecht University, Princetonplein 5, 3584
CC Utrecht, the Netherlands}

\author{N. Menezes}

\affiliation{Institute for Theoretical Physics, Centre for Extreme Matter and
Emergent Phenomena,\\ Utrecht University, Princetonplein 5, 3584
CC Utrecht, the Netherlands}

\author{Van Sérgio Alves}

\affiliation{Faculdade de Física, Universidade Federal do Pará,\\ Avenida Augusto
Correa 01, 66075-110 Belém, Pará, Brazil}

\author{C. Morais Smith}

\affiliation{Institute for Theoretical Physics, Centre for Extreme Matter and
Emergent Phenomena,\\ Utrecht University, Princetonplein 5, 3584
CC Utrecht, the Netherlands}

\date{\today}
\begin{abstract}
We use Pseudo Quantum Electrodynamics to study massive (2+1)D Dirac
systems interacting electromagnetically via a U(1) gauge field in
(3+1)D. It was recently found in Ref.~\cite{LeandroSilicene}, that
an interaction-induced Quantum Hall Effect (QHE) and Quantum Valley
Hall Effect (QVHE) occur in these systems, when considering a two-component
fermion representation. Here, we study the corrections to these effects
when coupling the fermions to a (2+1)D massive scalar field via a
quartic interaction. We find no correction to the QHE and a non-universal
correction to the QVHE, which depends on the ratio of the fermion
and scalar-field masses.
\end{abstract}
\maketitle

\section{Introduction}

With the experimental realization of graphene, a honeycomb lattice
of carbon atoms, massless 2D Dirac fermions moving with the Fermi
velocity $v_{\F}$ have been observed in condensed matter \cite{novoselov2005two}.
This discovery has triggered the application of the tools of relativistic
quantum-field theories in condensed-matter systems.

Initially, it was believed that the electrons in graphene were very
weakly interacting. However, the measurement of the fractional Quantum
Hall Effect \cite{FQHEgraphene} and of the renormalization of the
Fermi velocity \cite{FermiVrenormaliz,VozmedianoRG} have proven that
interactions are indeed important at low temperatures in sufficiently
clean samples. 

Considering static electron-electron interactions, both ac- \cite{jurivcic2010conductivity}
and dc-conductivity \cite{fritz2008quantum} have been calculated.
The possibility of a gap opening due to strong Coulomb or electron-phonon
interactions has also been investigated \cite{khveshchenko2009massive},
as well as the conductivity in the presence of both interactions and
an external magnetic field \cite{gusynin2006transport}. 

More recently, it was shown that dynamical electromagnetic interactions
may lead to a QVHE \cite{marino2015interaction}. In addition, the
correction to the bare spin $g$-factor due to dynamical interactions
has also been calculated \cite{menezes2016valley}, and were found
to exhibit good agreement with experiments \cite{kurganova2011spin,song2010high}
(see Ref.~\cite{kotov2012electron} for a comprehensive review of
electron-electron interactions in graphene).

Since the synthesis of graphene, many other 2D materials consisting
of a honeycomb lattice have been experimentally realized. One of these
is silicene, a honeycomb lattice made of silicon atoms \cite{kara2012review}.
The larger ionic radius of the silicon compared to carbon causes the
lattice to buckle and leads to a band-gap that can be tuned by a perpendicular
electric field. The low-energy excitations of silicene are thus \emph{massive}
Dirac fermions \cite{ezawa2012valley}. The buckled lattice also increases
the intrinsic spin-orbit coupling \cite{ezawa2012valley}. 

By describing silicene within a tight-binding Hamiltonian, including
spin-orbit coupling and a perpendicular electric field that explicitly
breaks the inversion symmetry, a non-universal QVHE was predicted
\cite{ValleyTopologicalSiliceneTahirM}. At the neutrality point,
however, the result becomes universal and depends only on the sign
of the spin-orbit and electric-field terms. Non-universal corrections
to the QVHE were also obtained in Ref.~\cite{ACDCspinValleyHall}
by including a finite chemical potential, and in Ref.~\cite{EzawaChernBrokenSymm}
by including a Rashba term that breaks the spin $s_{z}$-symmetry.

On the other hand, it was found in Ref.~\cite{LeandroSilicene} that
not only a QVHE, but also a QHE may emerge due to dynamical interactions
in massive Dirac systems, as a consequence of a dynamically driven
parity anomaly. In this case, the Hall (transverse) conductivity and
the Valley Hall conductivity assume universal values, depending only
the Planck constant $h$ and the electron charge $e$. It is remarkable
that the effect arises in the absence of a magnetic field or any other
perturbation that breaks time-reversal symmetry a priori.

Here, we investigate the fate of this universal QHE and QVHE when
we couple the system in Ref.~\cite{LeandroSilicene} to a (2+1)D
massive scalar field $\sigma$, via a quartic interaction. Scalar
fields have been used both in the context of electron-phonon interactions
and optomechanics to describe mechanical oscillations either of a
lattice \cite{roy2014migdal} or a movable mirror \cite{aspelmeyer2014cavity}.
Although the first term of the interaction in each of these systems
is linear in the scalar field, higher-order contributions can be considered.
For optomechanical systems, the quadratic term in the scalar field,
which generates the quartic interaction, would represent a quadratic
displacement of the oscillator's position \cite{aspelmeyer2014cavity}
and it has already been observed in a cold-atom setup \cite{purdy2010tunable}.
Moreover, this quartic coupling can be also found in a supersymmetric
generalization of Chern-Simons Higgs theory \cite{lee1990supersymmetry},
which was recently used in the non-relativistic limit to describe
the fractional quantum Hall effect \cite{tong2015quantum}.

We consider relativistic massive (2+1)D Dirac electrons, propagating
with a Fermi velocity $v_{\F}$ and interacting via a U(1) gauge field
that lives in (3+1)D. This dimensional mismatch is accounted for within
the framework of Pseudo Quantum Electrodynamics (PQED), the effective
theory that is obtained by integrating out the extra dimension of
the gauge field \cite{Marino1993}. The name Pseudo QED stems from
the fact that the theory involves pseudo-differential operators. This
theory is also sometimes called reduced QED in the literature \cite{gorbar2001dynamical,teber2014two,kotikov2014two}. 

Using the Kubo formalism, we obtain the correction to the transverse
conductivity induced by the coupling to the scalar field. We find
a non-universal correction to the QVHE, depending on the ratio of
the fermion and scalar-field masses, but no correction to the QHE. 

The outline of this paper is as follows: in Sec.~\ref{sec:The-Model}
we introduce the model. In Sec.~\ref{sec:Cur-cur}, we calculate
the current-current correlation function, which we use in Sec.~\ref{sec:Conductivity}
to obtain the correction to the conductivity. In Sec.~\ref{sec:Massless-case}
we consider the massless case and in Sec.~\ref{sec:Conclusions}
we present our conclusions. In the appendices we provide additional
details of our calculation.

\section{The Model\label{sec:The-Model}}

In 2D systems such as graphene and silicene, electrons interact via
a U(1) gauge field that propagates in (3+1)D. To describe this system,
one can start from QED in (3+1)D and confine the matter current $j^{\mu}$
to a plane \cite{Marino1993} by writing
\begin{align*}
j^{\mu}(x^{0},x^{1},x^{2},x^{3})=\begin{cases}
j_{2+1}^{\mu}(x^{0},x^{1},x^{2})\delta(x^{3}) & \mu=0,1,2\\
0 & \mu=3.
\end{cases}
\end{align*}
 The extra dimension of the gauge field can then be integrated out,
thus leading to a non-local theory, that is nevertheless causal \cite{do1992canonical}
and unitary \cite{marino2014unitarity}.

In this work, we start from PQED with massive, two component fermions
moving with a Fermi velocity $v_{\F}$. We couple the fermions to
a massive scalar field $\sigma$ via a quartic interaction. The Lagrangian
of the model reads
\begin{align}
\Lagr= & -\frac{1}{2}\frac{F^{\mu\nu}F_{\mu\nu}}{\sqrt{\square}}+\bar{\psi}_{a}\left(i\gamma^{0}\partial_{0}+iv_{\F}\gamma^{i}\partial_{i}-\Delta\right)\psi_{a}\nonumber \\
 & -e\bar{\psi}_{a}\gamma^{\mu}\psi_{a}A_{\mu}+\frac{1}{2}\partial^{\mu}\sigma\partial_{\mu}\sigma-\frac{1}{2}m_{\sigma}^{2}\sigma^{2}+g\bar{\psi}_{a}\psi_{a}\sigma^{2},\label{eq:Lagrangian}
\end{align}
where $\mu=0,1,2$, $F^{\mu\nu}$ is the electromagnetic tensor, $\psi$
is the electron field, $\Delta$ is the mass of the electron, $m_{\sigma}$
is the scalar-field mass, $g$ is the coupling constant of the quartic
interaction, $e$ is the electron charge, $A_{\mu}$ is the electromagnetic
4-potential, and $\gamma^{\mu}=\left(\gamma^{0},v_{\F}\gamma^{i}\right)$
are the gamma matrices. The electron field has a flavor index $a$
that specifies the valley and the spin component. We will consider
$N_{f}=4$, corresponding to two spin and two valley components. We
write the fermion mass as $\Delta=\xi m_{0}$, with the bare mass
$m_{0}>0$, $\xi=\pm1$ depending on the valley. In general, the mass
term breaks time-reversal symmetry, but since there are two valleys
connected by time-reversal conjugation, if the bare mass is $m_{0}$
for valley $K$ and $-m_{0}$ for valley $K'$, time-reversal symmetry
is preserved. We work in units where $\hbar=c=1$. Our model differs
from the one studied in Ref.~\cite{LeandroSilicene} because we add
a coupling between the fermions and a scalar field $\sigma$.

\section{Current-current correlation function\label{sec:Cur-cur}}

The conductivity can be calculated, in the linear-response regime,
using Kubo's formula
\begin{equation}
\sigma^{ij}=\lim_{\omega\rightarrow0,\mathbf{p}\rightarrow0}\frac{i\left\langle j^{i}j^{j}\right\rangle }{\omega}=\sigma_{xx}\delta^{ij}+\sigma_{xy}\epsilon^{ij},\label{eq:kuboFormula1}
\end{equation}
where $\left\langle j^{i}j^{k}\right\rangle $ is the current-current
correlation function, $\omega$ is the frequency, $\sigma_{xx}$ the
longitudinal and $\sigma_{xy}$ the transverse conductivity. The current-current
correlation function is nothing but the polarization tensor $\Pi^{ij}$.
Our strategy is to obtain the conductivity by computing the polarization
tensor, and then to apply Kubo's formula.

\begin{figure}
\input{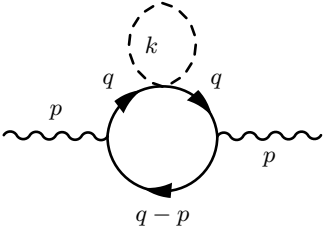}

\caption{2-loop diagram containing the $\sigma$ field contributing to $\Pi^{ij}.$\label{fig:2loopquartic}}

\end{figure}

We focus on the transverse conductivity, since the longitudinal conductivity
was shown to be zero for massive Dirac systems in the two-component
fermion representation, up to first order \cite{LeandroSilicene}.
We will calculate the lowest-order correction to the transverse part
of the vacuum polarization tensor coming from the scalar field $\sigma$,
to verify whether the coupling to the scalar field may destroy the
universal features of the transverse current. The lowest-order contribution
comes from the 2-loop diagram depicted in Fig.~\eqref{fig:2loopquartic}.
The corresponding expression is
\begin{align}
i\Pi_{2l}^{ij}(p,\Delta) & =2ie^{2}gv_{\mathrm{F}}^{2}\int\frac{d^{3}k}{(2\pi)^{3}}\frac{d^{3}q}{(2\pi)^{3}}\left\{ \frac{i}{k^{2}-m_{\sigma}^{2}}\right.\nonumber \\
 & \times\mathrm{Tr}\left[\gamma^{i}S_{F}(q)^{2}\gamma^{j}S_{F}(q-p)\right]\Biggr\},\label{eq:FullExpression}
\end{align}
where 
\[
S_{F}(q)=\frac{i\left(\gamma^{0}q_{0}+v_{\F}\gamma^{i}q_{i}+\Delta\right)}{q_{0}^{2}-v_{\F}^{2}\mathbf{q}^{2}-\Delta^{2}},
\]
is the fermion propagator. Note that there is a minus sign coming
from the fermionic loop, and a symmetry factor of two. We compute
the $k$-integral using dimensional regularization \cite{Dimreg1,Dimreg2,Dimreg3},
which yields
\begin{align}
\int\frac{d^{3}k}{(2\pi)^{3}}\frac{i}{k^{2}-m_{\sigma}^{2}} & =-\frac{m_{\sigma}}{4\pi}.\label{eq:LoopDiagram}
\end{align}
It is interesting to observe that using dimensional regularization,
we find a finite result. Since we have chosen the fermions to be two-component
spinors, the gamma matrices will also be two-dimensional. In this
representation, we can choose the gamma matrices such that they are
equal to the Pauli matrices, and we find 
\begin{equation}
\mathrm{Tr}\left[\gamma^{\mu}\gamma^{\nu}\gamma^{\rho}\right]=2i\epsilon^{\mu\nu\rho},\label{eq:GammaTrace}
\end{equation}
where $\epsilon^{\mu\nu\rho}$ is the Levi-Civita tensor. To find
the transverse conductivity, we have to identify the terms proportional
to $\epsilon^{ij}$. From Eq.~\eqref{eq:GammaTrace}, we see that
these terms will only arise from the trace of three and five gamma
matrices. Keeping only these terms (see appendix A for a detailed
calculation of the diagram), Eq.\eqref{eq:FullExpression} becomes
\begin{align}
i\Pi_{2l}^{ij} & =-ie{}^{2}gv_{\mathrm{F}}^{2}\frac{m_{\sigma}}{\pi}\nonumber \\
\times & \int\frac{d^{3}q}{(2\pi)^{3}}\left\{ \frac{\epsilon^{ij0}\left[\overline{q}^{2}(q_{0}-p_{0})-\Delta^{2}(q_{0}+p_{0})\right]}{\left(\overline{q}^{2}-\Delta^{2}\right)^{2}\left[(\overline{q}-\overline{p})^{2}-\Delta^{2}\right]}\right\} ,\label{eq:halfwayDiagram}
\end{align}
where we have introduced the notation $\overline{q}^{2}=q_{0}^{2}-v_{\F}^{2}\mathbf{q}^{2}$.
Next, we have to evaluate the $q$ integral. The result is 
\begin{align}
i\Pi_{2l}^{ij} & =-\frac{m_{\sigma}e^{2}g}{(4\pi)^{2}}i\epsilon^{ij0}p_{0}\nonumber \\
\times & \left[F_{1}(\Delta,p)+F_{2}(\Delta,p)\overline{p}^{2}+F_{3}(\Delta,p)\Delta^{2}\right],\label{eq:resultDiagram}
\end{align}
with 
\begin{align}
F_{1}(p,\Delta) & =-i\frac{20}{8}\frac{\left|\Delta\right|}{\overline{p}^{2}}-i\frac{\left(20\Delta^{2}-7\,\overline{p}^{2}\right)}{8\overline{p}^{3}}\ln\left[\frac{2\left|\Delta\right|-\left|\overline{p}\right|}{2\left|\Delta\right|+\left|\overline{p}\right|}\right],\nonumber \\
F_{2}(p,\Delta) & =\frac{-i\left|\Delta\right|}{4\Delta^{4}\overline{p}^{3}-\Delta^{2}\overline{p}^{5}}\Biggl\{2\left|\overline{p}\right|(2\Delta^{2}+\overline{p}^{2})\nonumber \\
 & \left.+\left|\Delta\right|(4\Delta^{2}-\overline{p}^{2})\ln\left[\frac{2\left|\Delta\right|-\left|\overline{p}\right|}{2\left|\Delta\right|+\left|\overline{p}\right|}\right]\right\} ,\nonumber \\
F_{3}(p,\Delta) & =\frac{-1}{8\overline{p}^{5}(-4\Delta^{2}+\overline{p}^{2})}\Biggl\{4i\left|\Delta\right|\overline{p}\left(-12\Delta^{2}+\overline{p}^{2}\right)\nonumber \\
 & -i\left(48\Delta^{4}-8\Delta^{2}\overline{p}^{2}-\overline{p}^{4}\right)\ln\left[\frac{2\left|\Delta\right|-\left|\overline{p}\right|}{2\left|\Delta\right|+\left|\overline{p}\right|}\right]\Biggr\}.\label{eq:formFactors}
\end{align}

\section{Conductivity\label{sec:Conductivity}}

We can now use the result obtained for the polarization tensor {[}Eq.~\eqref{eq:resultDiagram}{]}
to compute the corrections to the transverse conductivity due to the
scalar field. Because of the valley degree of freedom, there are two
valley currents in our model, which are connected by time-reversal
symmetry. From these two valley currents, we can define the total
conductivity
\begin{equation}
\sigma_{tot}^{ij}=\lim_{\omega\rightarrow0,\mathbf{p}\rightarrow0}\left\{ \frac{i\left\langle j^{i}j^{j}\right\rangle }{\omega}+\frac{i\left\langle j^{i}j^{j}\right\rangle ^{\mathrm{T}}}{\omega}\right\} =\sigma_{xx}^{tot}\delta^{ij}+\sigma_{xy}^{tot}\epsilon^{ij},\label{eq:TotalConductivity}
\end{equation}
and the valley conductivity
\begin{equation}
\sigma_{val}^{ij}=\lim_{\omega\rightarrow0,\mathbf{p}\rightarrow0}\left\{ \frac{i\left\langle j^{i}j^{j}\right\rangle }{\omega}-\frac{i\left\langle j^{i}j^{j}\right\rangle ^{\mathrm{T}}}{\omega}\right\} =\sigma_{xx}^{val}\delta^{ij}+\sigma_{xy}^{val}\epsilon^{ij},\label{eq:ValleyConductivity}
\end{equation}
 where $\left\langle j^{i}j^{j}\right\rangle ^{\mathrm{T}}$ is the
time-reversed current-current correlation function. In Ref.~\cite{LeandroSilicene},
it was found that 
\begin{align}
\sigma_{xy}^{val} & =4\left(n+\frac{1}{2}\right)\frac{e^{2}}{h},\label{eq:Leandrosresults1}\\
\sigma_{xy}^{tot} & =2\frac{e^{2}}{h}.\label{eq:leandroResults}
\end{align}
 From the polarization tensor in Eq.~\eqref{eq:resultDiagram}, we
find the correction to the current 
\begin{align}
\lim_{\omega\rightarrow0,\mathbf{p}\rightarrow0}\frac{i\left\langle j^{i}j^{j}\right\rangle }{\omega} & =\lim_{\omega\rightarrow0,\mathbf{p}\rightarrow0}\frac{i\Pi_{2-loop}^{ij}(p,\Delta)}{\omega},\label{eq:kubolim1}\\
\lim_{\omega\rightarrow0,\mathbf{p}\rightarrow0}\frac{i\left\langle j^{i}j^{j}\right\rangle ^{\mathrm{T}}}{\omega} & =\lim_{\omega\rightarrow0,\mathbf{p}\rightarrow0}\frac{\left[\Pi_{2-loop}^{ij}(p,\Delta)\right]^{\mathrm{T}}}{\omega},\label{eq:kubolim2}
\end{align}
where we recall that $p_{0}=\omega$. At first glance, it seems that
the expressions in Eq.~\eqref{eq:formFactors} are not well defined
in the Kubo limit. However, when taking all the terms together and
considering the Taylor expansion of the logarithms for small $p_{0}$,
we find that the divergences cancel (see appendix B for details).
We find
\begin{align}
\lim_{\omega\rightarrow0,\mathbf{p}\rightarrow0}F_{1}(p,\Delta) & =-\frac{2}{3}\frac{i}{\left|\Delta\right|},\label{eq:f1lim}\\
\lim_{\omega\rightarrow0,\mathbf{p}\rightarrow0}F_{2}(\Delta,p)\overline{p}^{2} & =0,\label{eq:f2lim}\\
\lim_{\omega\rightarrow0,\mathbf{p}\rightarrow0}F_{3}(\Delta,p)\Delta^{2} & =-\frac{1}{30}\frac{i}{\left|\Delta\right|}.\label{eq:f3lim}
\end{align}

Finally, after substituting Eqs.~\eqref{eq:f1lim}-\eqref{eq:f3lim}
into Eq.~\eqref{eq:resultDiagram}, and this into Eqs.~\eqref{eq:kubolim1}
and \eqref{eq:kubolim2}, taking into account a factor of 2 for the
spin degree of freedom, and reintroducing $\hbar$ to make the result
dimensional, we find a non-universal correction to the valley conductivity
\begin{align}
\delta\sigma_{xy}^{val}= & -\frac{1}{\left(2\pi\right)^{2}}\frac{7}{10}\frac{m_{\sigma}}{\left|\Delta\right|}g\frac{e^{2}}{h},\label{eq:valcorrection}
\end{align}
and no correction to the total conductivity 
\begin{align}
\delta\sigma_{xy}^{tot}= & 0.\label{eq:totcorrection}
\end{align}
Combining this result with the result from Ref.~\cite{LeandroSilicene}
{[}Eqs.~\eqref{eq:Leandrosresults1} and \eqref{eq:leandroResults}{]},
we find that 
\begin{align}
\sigma_{xy}^{val} & =2\frac{e^{2}}{h}\left(2n+1-\frac{1}{\left(2\pi\right)^{2}}\frac{7}{20}g\frac{m_{\sigma}}{\left|\Delta\right|}\right),\nonumber \\
\sigma_{xy}^{tot} & =2\frac{e^{2}}{h}.\label{eq:CombinedResult}
\end{align}

\section{The massless case\label{sec:Massless-case}}

Let us now investigate the correction to the polarization tensor for
massless fermions (as found, for example, in graphene). In the massless
case $\Delta=0$, Eq.~\eqref{eq:halfwayDiagram} then reduces to
\begin{align}
i\Pi_{2l}^{ij} & =-ie{}^{2}gv_{\mathrm{F}}^{2}\frac{m_{\sigma}}{\pi}\int\frac{d^{3}q}{(2\pi)^{3}}\left\{ \frac{\epsilon^{ij0}(q_{0}-p_{0})}{\overline{q}^{2}(\overline{q}-\overline{p})^{2}}\right\} .\label{eq:MassLessDiagram}
\end{align}
Combining the denominators and calculating the integrals, we find
\begin{align}
i\Pi_{2l}^{ij} & =\frac{m_{\sigma}e{}^{2}g}{16\pi}i\epsilon^{ij0}\frac{p_{0}}{\sqrt{p_{0}^{2}-v_{\F}^{2}\mathbf{p}^{2}}}.\label{eq:MasslessResult}
\end{align}
In this case, the Kubo formula is not well defined since dividing
by $\omega$ and taking the zero limit of the momentum yields a divergence.
We could already expect this result on dimensional grounds. The polarization
tensor in our theory has mass dimension one. Integrating out the bosonic
loop, we find something proportional to the mass $m_{\sigma}$. The
final result should thus be $m_{\sigma}$ multiplied by a dimensionless
term. We also know that the transverse part will be proportional to
$\epsilon^{ij}p_{0}$, and thus we need to divide by a term with mass
dimension one in order to make the Kubo formula well-defined. In the
massless case, we can only divide by a term containing $p_{0}$ and
$v_{\F}\mathbf{p}$. This means that the Kubo limit will not be well-defined.
When we have a fermion mass, we can also divide by this mass to make
the limit finite, and indeed this is exactly what happens in Eqs.~\eqref{eq:kubolim1}
and \eqref{eq:kubolim2}.

\section{Conclusion\label{sec:Conclusions}}

It has been known for some time that for QED in (2+1)D, in the two-component
spinor representation, radiative corrections generate a topological
gauge field mass term \cite{TopoMass1,TopoMass2}, giving rise to
a non-vanishing transverse current in the system \cite{niemi1983axial,khalilov2000polarization}.
Although this induced mass term emerges when one couples the fermions
minimally to the vector potential $A_{\mu}$, interactions between
fermions and other fields could lead to additional contributions to
the current. Recently, it was shown that dynamical interactions described
within the PQED formalism lead to quantized Hall and valley Hall conductivities
\cite{LeandroSilicene}. At one-loop order, the results for QED and
PQED in (2+1)D are the same. At higher order, however, they differ
for the longitudinal conductivity, but remain the same for the transverse
one \cite{coleman1985no}. Here, we investigated the fate of these
quantized conductivities in the presence of an additional scalar field
quartically coupled to the fermions. 

We started by calculating the corrections to the interaction induced
QVHE and QHE in massive Dirac systems using the PQED formalism, which
takes into account the full dynamical electromagnetic interactions
of the electrons. The corrections to the transverse conductivity and
transverse valley conductivity were obtained by calculating the polarization
tensor diagram up to 2-loop orders, and then using the Kubo formula.
We found a non-universal correction to the QVHE, which depends on
the ratio of the masses of the scalar field and the fermions, but
no correction to the QHE. In addition, we investigated the case of
massless fermions ($\Delta=0$), and showed that the Kubo formula
is not well defined in this limit. In the case of massless bosons
($m_{\sigma}=0$), there is no correction to either the QVHE or the
QHE. 

Here, we considered a not so explored quartic coupling between the
scalar and the fermionic fields. A Yukawa-like coupling was used in
the context of electron-phonon interaction in graphene \cite{YukawaBitenRoy}.
A theory involving an exponential of a scalar field was recently proposed
to describe fractionalization in a square lattice \cite{seradjeh2008fractionalization}.
A second-order expansion of an exponential containing a scalar field
would inevitably lead to a theory involving a Yukawa term plus the
quartic interaction considered here. We hope that our paper will motivate
further research on these non-standard couplings.
\begin{acknowledgments}
The authors would like to thank Giandomenico Palumbo, Leandro O. Nascimento,
Eduardo Marino, Guido van Miert and Anton Quelle for fruitful discussions.
S.H. Kooi is grateful to the Netherlands Organisation for Scientific
Research (NWO) for financial support. This work was supported by the
CNPq through the Brazilian government project Science Without Borders.
\end{acknowledgments}

\section*{Appendix A: calculation of the diagram\label{sec:Appendix-A}}

In this appendix, we show the detailed calculation of the correction
to the polarization tensor. Because Lorentz invariance is broken by
the Fermi velocity $v_{\F}$ of the fermions, we must treat the $q_{0}$
and $\mathbf{q}$ integrals separately. Let us start from Eq.~\eqref{eq:FullExpression},
and first compute the trace of the gamma matrices. As explained in
the main text, we are interested in the terms proportional to $\epsilon^{ij0}p_{0}$,
which can only arise from the trace of three or five gamma matrices.
The full trace is (we use the notation $\overline{q}=(q_{0},v_{\F}\boldsymbol{q})$)
\[
\mathrm{Tr}\left[\gamma^{i}\left(\gamma^{\alpha}\overline{q}_{\alpha}+\Delta\right){}^{2}\gamma^{j}\left(\gamma^{\beta}(\overline{q}-\overline{p})_{\beta}+\Delta\right)\right].
\]
The terms with three gamma matrices are 
\begin{align*}
 & 2\Delta^{2}\Tr\left[\gamma^{i}\gamma^{0}\gamma^{j}\right]q_{0}+\Delta^{2}\Tr\left[\gamma^{i}\gamma^{j}\gamma^{0}\right]\left(q_{0}-p_{0}\right)\\
= & 2\Delta^{2}\Tr\left[\gamma^{i}\gamma^{0}\gamma^{j}\right]q_{0}-\Delta^{2}\Tr\left[\gamma^{i}\gamma^{0}\gamma^{j}\right]\left(q_{0}-p_{0}\right)\\
= & \Delta^{2}\Tr\left[\gamma^{i}\gamma^{0}\gamma^{j}\right](q_{0}+p_{0})\\
= & 2\Delta^{2}i\epsilon^{i0j}\left(q_{0}+p_{0}\right),
\end{align*}
where we have used Eq.~\eqref{eq:GammaTrace}. There is one term
containing five gamma matrices, which is
\begin{align*}
 & \Tr\left[\gamma^{i}\gamma^{\alpha}\gamma^{\beta}\gamma^{j}\gamma^{\delta}\right]\overline{q}_{\alpha}\overline{q}_{\beta}\left(\overline{q}-\overline{p}\right)_{\delta}\\
= & \left\{ -\Tr\left[\gamma^{i}\gamma^{\beta}\gamma^{\alpha}\gamma^{j}\gamma^{\delta}\right]+2g^{\alpha\beta}\Tr\left[\gamma^{i}\gamma^{j}\gamma^{\delta}\right]\right\} \overline{q}_{\alpha}\overline{q}_{\beta}\left(\overline{q}-\overline{p}\right)_{\delta}\\
= & -\Tr\left[\gamma^{i}\gamma^{\alpha}\gamma^{\beta}\gamma^{j}\gamma^{\delta}\right]\overline{q}_{\alpha}\overline{q}_{\beta}\left(\overline{q}-\overline{p}\right)_{\delta}+4i\epsilon^{ij0}\overline{q}^{2}(q_{0}-p_{0}),
\end{align*}
from which it follows 
\[
\Tr\left[\gamma^{i}\gamma^{\alpha}\gamma^{\beta}\gamma^{j}\gamma^{\delta}\right]\overline{q}_{\alpha}\overline{q}_{\beta}\left(\overline{q}-\overline{p}\right)_{\delta}=2i\epsilon^{ij0}\overline{q}^{2}\left(q_{0}-p_{0}\right).
\]
Substituting the result for the trace in Eq.~\eqref{eq:FullExpression},
we obtain Eq.~\eqref{eq:halfwayDiagram}. We now combine the denominators
using the Feynman trick
\begin{align}
\frac{1}{A^{2}B} & =2\intop_{0}^{1}dx\frac{(1-x)}{\left[(1-x)A+xB\right]^{3}}.\label{eq:FeynmanTrick}
\end{align}
The denominator becomes
\begin{align}
(1-x)A+xB & =\left(q_{0}-xp_{0}\right)^{2}-\Sigma_{1},\label{eq:DenominatorCombined}
\end{align}
with 
\[
\Sigma_{1}\equiv-x(1-x)p_{0}^{2}+v_{\F}^{2}\mathbf{q}^{2}+\Delta^{2}+x\left[v_{\F}^{2}\mathbf{p}^{2}-2v_{\F}^{2}\mathbf{pq}\right].
\]
 Rewriting Eq.~\eqref{eq:halfwayDiagram} using Eqs.~\eqref{eq:FeynmanTrick}
and \eqref{eq:DenominatorCombined}, then making the shift $q_{0}\rightarrow q_{0}+xp_{0}$,
and noticing that the terms odd in $q_{0}$ vanish, the polarization
tensor becomes
\begin{align}
i\Pi_{2l}^{ij} & =-ie{}^{2}gv_{\mathrm{F}}^{2}\frac{m_{\sigma}}{\pi}2\intop_{0}^{1}dx\int\frac{d^{3}q}{(2\pi)^{3}}\nonumber \\
 & \times\left\{ \frac{\epsilon^{ij0}\left(q_{0}^{2}C+D\right)(1-x)}{\left(q_{0}^{2}-\Sigma_{1}\right)^{3}}\right\} ,
\end{align}
with 
\begin{align*}
C & \equiv(x-1)p_{0}+2xp_{0},\\
D & \equiv x^{2}(x-1)p_{0}^{3}-v_{\F}^{2}\mathbf{q}^{2}(x-1)p_{0}-\Delta^{2}(1+x)p_{0}.
\end{align*}
We can now perform the $q_{0}$ integrals
\begin{align*}
\int\frac{dq_{0}}{(2\pi)}\frac{q_{0}^{2}}{\left(q_{0}^{2}-\Sigma_{1}\right)^{3}} & =\frac{-i}{16}\Sigma_{1}^{-3/2},\\
\int\frac{dq_{0}}{(2\pi)}\frac{1}{\left(q_{0}^{2}-\Sigma_{1}\right)^{3}} & =i\frac{3}{16}\Sigma_{1}^{-5/2}.
\end{align*}
Rewriting 
\begin{align}
\Sigma_{1} & =v_{\F}^{2}\left[\left(\mathbf{q}-xp\right)^{2}-\Sigma_{2}\right],\label{eq:SigmaRewrite}
\end{align}
with 
\[
\Sigma_{2}\equiv-x(1-x)\mathbf{p}^{2}-\frac{\Delta^{2}}{v_{\F}^{2}}+x(1-x)p_{0}^{2}\frac{1}{v_{\F}^{2}},
\]
we find
\begin{align}
i\Pi_{2l}^{ij} & =-ie{}^{2}gv_{\mathrm{F}}^{2}\frac{m_{\sigma}}{8\pi}\intop_{0}^{1}dx\int\frac{d^{2}\mathbf{q}}{(2\pi)^{2}}\nonumber \\
\times & \left\{ \frac{-i\epsilon^{ij0}C(1-x)}{v_{\F}^{3}\left[\left(\mathbf{q}-x\mathbf{p}\right)^{2}-\Sigma_{2}\right]^{3/2}}+\frac{3i\epsilon^{ij0}D(1-x)}{v_{\F}^{5}\left[\left(\mathbf{q}-x\mathbf{p}\right)^{2}-\Sigma_{2}\right]^{5/2}}\right\} .\label{eq:AfterQ0Integral}
\end{align}
We now shift $\mathbf{q}\rightarrow\mathbf{q}+x\mathbf{p}$ and notice
that the terms odd in $\mathbf{q}$ vanish. The $\mathbf{q}$ integrals
may be performed using
\begin{align*}
\int\frac{d^{2}\mathbf{q}}{(2\pi)^{2}}\frac{1}{\left(\mathbf{q}^{2}-\Sigma_{2}\right)^{3/2}} & =\frac{-i}{2\pi}\frac{1}{\sqrt{\Sigma_{2}}},\\
\int\frac{d^{2}\mathbf{q}}{(2\pi)^{2}}\frac{\mathbf{q}^{2}}{\left(\mathbf{q}^{2}-\Sigma_{2}\right)^{5/2}} & =\frac{-i}{2\pi}\frac{2}{3}\frac{1}{\sqrt{\Sigma_{2}}},\\
\int\frac{d^{2}\mathbf{q}}{(2\pi)^{2}}\frac{1}{\left(\mathbf{q}^{2}-\Sigma_{2}\right)^{5/2}} & =\frac{i}{2\pi}\frac{1}{3}\frac{1}{\left(\Sigma_{2}\right)^{3/2}}.
\end{align*}
The polarization tensor then becomes 
\begin{align}
i\Pi_{2l}^{ij} & =-ie{}^{2}gv_{\mathrm{F}}^{2}\frac{m_{\sigma}}{16\pi^{2}}\intop_{0}^{1}dx\nonumber \\
\times & \left\{ \underbrace{\frac{\epsilon^{ij0}\left[-C-2(x-1)p_{0}\right](1-x)}{v_{\F}^{3}\left(\Sigma_{2}\right)^{1/2}}}_{I_{1}}\underbrace{-\frac{\epsilon^{ij0}E(1-x)}{v_{\F}^{5}\left(\Sigma_{2}\right)^{3/2}}}_{I_{2}}\right\} ,\label{eq:DiagramAlmostFinal}
\end{align}
 with $E\equiv x^{2}(x-1)p_{0}^{3}-v_{\F}^{2}x^{2}\mathbf{p}^{2}(x-1)p_{0}-\Delta^{2}(1+x)p_{0}$. 

We are now left with only the parametric integral over $x$. The first
term of the integral is
\begin{align}
I_{1} & =\epsilon^{ij0}p_{0}\frac{1}{v_{\F}^{2}}\intop_{0}^{1}dx\frac{(3-5x)(1-x)}{\left[x(1-x)\overline{p}^{2}-\Delta^{2}\right]^{1/2}}\nonumber \\
 & =\epsilon^{ij0}p_{0}\frac{1}{v_{\F}^{2}}\Biggl\{-i\frac{20}{8}\frac{\left|\Delta\right|}{\overline{p}^{2}}\nonumber \\
 & -i\frac{\left(20\Delta^{2}-7\overline{p}^{2}\right)}{8\overline{p}^{3}}\ln\left(\frac{2\left|\Delta\right|-\left|\overline{p}\right|}{2\left|\Delta\right|+\left|\overline{p}\right|}\right)\Biggr\}\nonumber \\
 & =\epsilon^{ij0}p_{0}\frac{1}{v_{\F}^{2}}F_{1}(\Delta,p),\label{eq:int1}
\end{align}
and the second term is 
\begin{align}
I_{2} & =\epsilon^{ij0}\frac{1}{v_{\F}^{2}}\intop_{0}^{1}dx\frac{\left[\overbrace{-x^{2}(1-x)^{2}\overline{p}^{2}p_{0}}^{I_{2A}}-\overbrace{\Delta^{2}(1-x^{2})p_{0}}^{I_{2B}}\right]}{\left[x(1-x)\overline{p}^{2}-\Delta^{2}\right]^{3/2}},\label{eq:int2}
\end{align}
where 
\begin{align}
I_{2A} & =\intop_{0}^{1}dx\frac{-x^{2}(1-x)^{2}\overline{p}^{2}p_{0}}{\left[x(1-x)\overline{p}^{2}-\Delta^{2}\right]^{3/2}}\nonumber \\
 & =\frac{-1}{4\Delta^{4}\overline{p}^{3}-\Delta^{2}\overline{p}^{5}}\Biggl\{2i\left|\Delta\right|\left|\overline{p}\right|(2\Delta^{2}+\overline{p}^{2})\nonumber \\
 & \left.+i\Delta^{2}(4\Delta^{2}-\overline{p}^{2})\ln\left[\frac{2\left|\Delta\right|-\left|\overline{p}\right|}{2\left|\Delta\right|+\left|\overline{p}\right|}\right]\overline{p}^{2}p_{0}\right\} \nonumber \\
 & =F_{2A}(\Delta,p)\overline{p}^{2}p_{0},\label{eq:int2A}
\end{align}
and

\begin{align}
I_{2B} & =\intop_{0}^{1}dx\frac{-(1-x^{2})\Delta^{2}p_{0}}{\left[x(1-x)\overline{p}^{2}-\Delta^{2}\right]^{3/2}}\nonumber \\
 & =\frac{-1}{8\overline{p}^{5}(-4\Delta^{2}+\overline{p}^{2})}\Biggl\{4i\left|\Delta\right|\overline{p}\left(-12\Delta^{2}+\overline{p}^{2}\right)\nonumber \\
 & -i\left(48\Delta^{4}-8\Delta^{2}\overline{p}^{2}-\overline{p}^{4}\right)\ln\left[\frac{2\left|\Delta\right|-\left|\overline{p}\right|}{2\left|\Delta\right|+\left|\overline{p}\right|}\right]\Biggr\}\Delta^{2}p_{0}\nonumber \\
 & =F_{2B}(\Delta,p)\Delta^{2}p_{0}.\label{eq:int2B}
\end{align}
Substituting Eqs.~\eqref{eq:int1}-\eqref{eq:int2B} into Eq.~\eqref{eq:DiagramAlmostFinal}
leads to Eq.~\eqref{eq:resultDiagram} in the main text.

\section*{Appendix B: Taking the Kubo limit\label{sec:Appendix-B}}

In this section we calculate the Kubo limit of Eq.\eqref{eq:resultDiagram}:
\begin{align}
\lim_{p_{0}\rightarrow0,\mathbf{p\rightarrow0}}\frac{\Pi_{2l}^{ij}}{p_{0}} & =\lim_{p_{0}\rightarrow0,\mathbf{p\rightarrow0}}-\frac{m_{\sigma}e^{2}g}{(4\pi)^{2}}\epsilon^{ij0}\nonumber \\
 & \times\left[F_{1}(\Delta,p)+F_{2}(\Delta,p)\overline{p}^{2}+F_{3}(\Delta,p)\Delta^{2}\right],\label{eq:kubolimitapp}
\end{align}
where the expressions for $F_{1}(\Delta,p)$, $F_{2}(\Delta,p)\overline{p}^{2}$
and $F_{3}(\Delta,p)\Delta^{2}$ are given in Eq.~\eqref{eq:formFactors}.
We first note that Eq.\eqref{eq:kubolimitapp} is only dependent on
$\left|\overline{p}\right|$, and taking the limit $\mathbf{p}\rightarrow0$,
thus amounts to replacing $\left|\overline{p}\right|\rightarrow\left|p_{0}\right|$.
To take the limit of $p_{0}\rightarrow0$ we have to consider the
Taylor expansion
\begin{align*}
\ln\left[\frac{2\left|\Delta\right|-\left|p_{0}\right|}{2\left|\Delta\right|+\left|p_{0}\right|}\right] & =-\frac{\left|p_{0}\right|}{\left|\Delta\right|}-\frac{1}{12}\frac{\left|p_{0}\right|^{3}}{\left|\Delta\right|^{3}}-\frac{1}{80}\frac{\left|p_{0}\right|^{5}}{\left|\Delta\right|^{5}}+O\left(p_{0}^{6}\right).
\end{align*}
We calculate the limit for each term separately. For the first term,
we find
\begin{align*}
 & \lim_{p_{0}\rightarrow0}F_{1}(\Delta,p_{0})\\
= & \lim_{p_{0}\rightarrow0}-i\frac{20}{8}\frac{\left|\Delta\right|}{p_{0}^{2}}-i\frac{\left(20\Delta^{2}-7\,p_{0}^{2}\right)}{8\left|p_{0}\right|^{3}}\ln\left[\frac{2\left|\Delta\right|-\left|p_{0}\right|}{2\left|\Delta\right|+\left|p_{0}\right|}\right]\\
= & \lim_{p_{0}\rightarrow0}-i\frac{20}{8}\frac{\left|\Delta\right|}{p_{0}^{2}}-i\frac{\left(20\left|\Delta\right|^{2}-7p_{0}^{2}\right)}{8\left|p_{0}\right|^{3}}\left(-\frac{\left|p_{0}\right|}{\left|\Delta\right|}-\frac{1}{12}\frac{\left|p_{0}\right|^{3}}{\left|\Delta\right|^{3}}\right)\\
 & +O(p_{0})\\
= & \lim_{p_{0}\rightarrow0}-\frac{16}{24}\frac{i}{\left|\Delta\right|}\\
= & -\frac{2}{3}\frac{i}{\left|\Delta\right|}.
\end{align*}
The second term becomes
\begin{align*}
 & \lim_{p_{0}\rightarrow0}F_{2}(\Delta,p_{0})p_{0}^{2}\\
= & \lim_{p_{0}\rightarrow0}\frac{-i\left|\Delta\right|p_{0}^{2}}{4\Delta^{4}\left|p_{0}\right|^{3}-\Delta^{2}\left|p_{0}\right|^{5}}\Biggl\{2\left|p_{0}\right|(2\Delta^{2}+p_{0}^{2})\\
 & +\left|\Delta\right|(4\Delta^{2}-p_{0}^{2})\ln\left[\frac{2\left|\Delta\right|-\left|p_{0}\right|}{2\left|\Delta\right|+\left|p_{0}\right|}\right]\Biggr\}\\
= & \lim_{p_{0}\rightarrow0}\frac{-4i\left|\Delta\right|}{\left(4\Delta^{2}-p_{0}^{2}\right)}-\frac{2ip_{0}^{2}}{\left|\Delta\right|\left(4\Delta^{2}-p_{0}^{2}\right)}-\frac{i}{\left|p_{0}\right|}\left(-\frac{\left|p_{0}\right|}{\left|\Delta\right|}\right)\\
 & +O(p_{0})\\
= & \frac{-4i\left|\Delta\right|}{4\left|\Delta\right|^{2}}+\frac{i}{\left|\Delta\right|}\\
= & 0,
\end{align*}
and the third term becomes
\begin{align*}
 & \lim_{p_{0}\rightarrow0}F_{3}(\Delta,p_{0})\Delta^{2}\\
= & \lim_{p_{0}\rightarrow0}\frac{-\Delta^{2}}{8\overline{p}^{5}(-4\Delta^{2}+p_{0}^{2})}\Biggl\{4i\left|\Delta\right|p_{0}\left(-12\Delta^{2}+p_{0}^{2}\right)\\
 & -i\left(48\Delta^{4}-8\Delta^{2}p_{0}^{2}-p_{0}^{4}\right)\ln\left[\frac{2\left|\Delta\right|-\left|p_{0}\right|}{2\left|\Delta\right|+\left|p_{0}\right|}\right]\Biggr\}.\\
= & \lim_{p_{0}\rightarrow0}\frac{i\left|\Delta\right|}{12(p_{0}^{2}-4\left|\Delta\right|^{2})}+\frac{i\left|\Delta\right|}{8(p_{0}^{2}-4\left|\Delta\right|^{2})}\\
 & -\frac{i6\left|\Delta\right|}{80(p_{0}^{2}-4\left|\Delta\right|^{2})}+O(p_{0})\\
= & -\frac{1}{30}\frac{i}{\left|\Delta\right|}.
\end{align*}

\bibliographystyle{apsrev4-1}
\bibliography{references}

\end{document}